# Reflected Light from Sand Grains in the Terrestrial Zone of a Protoplanetary Disk

William Herbst[*], Catrina M. Hamilton[†], Katherine LeDuc[*], Joshua N. Winn[‡], Christopher M. Johns-Krull[§], Reinhard Mundt[|], Mansur Ibrahimov[¶]

[*]*Astronomy Dept., Wesleyan University, Middletown, CT 06459, USA*

[†]*Physics and Astronomy Dept., Dickinson College, Carlisle, PA 17013, USA*

[‡]*Dept. of Physics, Massachusetts Institute of Technology, 77 Massachusetts Ave., Cambridge, MA 02139 USA*

[§]*Dept. of Physics and Astronomy, Rice University, Houston, TX 77005*

[|]*Max-Planck-Institut fürAstronomie, Königstuhl 17, D-69117 Heidelberg, Germany*

[¶]*Ulugh Beg Astronomical Institute of the Uzbek Academy of Sciences, Astronomicheskaya 33, 700052 Tashkent, Uzbekistan*

**In the standard model of terrestrial planet formation, the first step is for interstellar dust to coagulate within a protoplanetary disk surrounding a young star, forming large grains that settle towards the disk plane.[1] Interstellar grains of typical size ~0.1 μ are expected to grow to millimetre (sand), centimeter (pebble) or even meter-sized (boulder) objects rather quickly.[2] Unfortunately, such evolved disks are hard to observe because of the small surface area to volume ratio of their constituents. While we readily detect dust around young objects known as "classical" T Tauri stars, there is little or no evidence of it in the slightly more evolved "weak-line" systems.[3] Here we show that grains have grown to ~mm size**

**or larger in the terrestrial zone (within ~3 AU) of a 3 Myr old star. The fortuitous geometry of the KH 15D binary star system allows us to infer that, when faint, it appears as a nearly edge-on ring front-illuminated by one of the central binary components. This work complements the study of terrestrial zones of younger disks that have recently been resolved by interferometry[4] and affords an opportunity to determine optical and physical properties of evolved grains in a protoplanetary disk.**

Fig. 1 shows two views of the geometry of the KH 15D system according to our present interpretation. We adopt an updated version of Model 3 of Winn et al.[5] for all quantitative discussion. KH 15D is known to be a pre-main sequence binary system with an age of ~3 Myr and a distance of 760 pc.[6,7,8] It consists of 0.6 and 0.7 $M_\odot$ stars (designated A and B, respectively[5]) in an eccentric orbit (e=0.6) with a period of 48.37 days. Its orbit is inclined to a circumstellar disk that extends to ~5 AU[9]; precession of the disk is causing an occulting edge to advance across the orbit of the binary, as shown in Fig. 1.[9,10] The precession time-scale for the disk is ~1000 yrs and it must either be warped[9] or flared (or both) to cause both the foreground occultation and the background reflection. It is not known where, within this ring, the occulting edge is actually located – it could be at the inner or outer edge or at the high point of a warp somewhere in between. See Fig. 6 of the model paper[5] for a sketch of the possibilities.

For the past decade, only one star (A) has appeared above the disk edge, undergoing a dramatic "sunrise" and "sunset" every orbital cycle. During this time interval, the brightness of the system as measured at Earth has depended primarily on only one variable, namely the elevation (Δ) of star A above (or below) the "horizon" defined by the edge of the occulting screen as projected on the plane of the sky. We measure Δ in units of the radius of star A (1.3 $R_\odot \approx 10^9$ m). Fig. 2 shows the system brightness at 0.80 μ, as a function of Δ, during the last four years.[11,12] For Δ>1 the full disk of star A is visible and the system's flux is determined by its brightness; for +1>Δ>-1 the star is partly occulted, and for Δ<-1 it is fully occulted. As the inset shows, while the total light from the KH 15D system plummets precipitously in the interval +1>Δ>-1, it also continues to depend on Δ even after full occultation (Δ<-1).

In earlier studies, the system's visibility after full occultation has been attributed to halos of scattered light around each component or around the system as a whole.[5,13] Analysis of high-resolution spectra of the system when faint, now paints a different picture. Fig. 3 displays the cross-correlation functions of a number of high-resolution spectra with a K7 V synthetic spectrum. The spectra show a single strong peak at the radial velocity of star A for Δ>-1, but become double-peaked after full occultation, with one peak corresponding to star A and the other to a red-shifted component. The radial velocity of the second component does not match star B, which is also far below the horizon at these times (see Fig. 1) and heavily obscured. We propose that the second peak is reflected light from the part of KH 15D's disk that is behind the stars from our vantage point. Star A is moving towards the Earth (relative to the systemic velocity of +18.6 km/s) during both ingress and egress so its reflection from a mirror behind the star would appear to be moving away from us (red-shifted). During mid-eclipse, it is star B that is more elevated with respect to the obscuring edge and we may see its direct light (very dimly) and its reflection, as the upper right panel in Fig. 3 shows. Clearly, there is excellent agreement, especially during ingress, between the predicted radial velocities according to this interpretation and the observed velocities of the red-shifted components.

Our reflection model also accounts well for the shape of the light variations during full occultation, as we now discuss (see Supplement for additional detail). The inclination of the orbit to the ring means that, viewed from a point on the "top surface" (i.e. the surface visible from Earth) of the back side of the ring, only one star is fully visible at any time and its height above the local horizon depends on the orbital phase. We assume that the brightness of the system is proportional to the solid angle subtended by the top of the ring as seen from the illuminating star. We also allow for a transmission component to the light, as required by the spectra. From simultaneous photometry obtained on three nights when we could separate the direct and reflected light, we can estimate the optical depth of the occulting screen as a function of $\Delta$. These data show that $\tau$=2.9 at $\Delta$=-1 and that it increases roughly linearly with $\Delta$, as one would expect. The reflection is assumed to come from a flat ring with inner and outer radii of 0.5 and 5 AU; a representative model light curve is shown as the solid line in the inset of Fig. 2. Clearly, it reproduces the shape of the observed light curve quite well, in particular the brightening that occurs for $\Delta$<-6 due to the fact that star B has risen to become the object's source of illumination.

The data and our interpretation also permit us to derive constraints on the size of the grains that constitute the obscuring/reflecting disk. Fig. 4 shows how the colour of the system changes during full occultation in the optical and near-infrared. Importantly there is no evidence for reddening, which implies that the ring particles are significantly larger (by at least a factor of 20) than those in the interstellar medium. Immediately after star A "sets" ($\Delta$<-1), some light from that star is still visible through the occulting part of the disk allowing a measurement of the optical depth along that path length, which provides a further size constraint on the grains. In our model, the disk optical thickness ($\tau$), can be expressed as $\tau = 64MR^{-2}r^{-1}\rho^{-1}(L/T)$, where M is the total mass of obscuring particles (in Earth masses), R is the disk radius (in AU), r is the grain radius (in mm), $\rho$ is the density of a grain (in $gm/cm^3$), L is the path length of the radiation through the obscuring medium and T is the disk thickness. Our data show that $\tau$=2.9 at $\Delta$=-1, so for representative choices of $\rho$=3 and L/T=0.05 we obtain $r \cong \Sigma$, where $\Sigma$ is the mass surface density of the disk in Earth masses per square AU. Using our own solar system as a guide, we may infer a grain size for this path length through the disk of roughly 1 mm. Interestingly, this is characteristic of chondrules, the glassy spherules that are a primary component of the most primitive meteorites found in our own solar system.[22]

The most uncertain parameter in our expression for $\tau$ is (L/T) and this could be larger than the value of 0.05 we adopt, which would increase the inferred size of the grains. Our derivation (see Supplementary Material) assumes that the grains are uniformly distributed through a cylindrical volume of thickness T, where T is determined from the projected extent of the obscuring material on the plane of the sky. A more realistic picture may be that the grains are concentrated into a thinner structure, which is tilted, warped, corrugated and/or flared in such a way that its footprint of obscuration on the sky is much larger than its physical thickness. In this case L/T > 0.05, and one might have a disk composed of pebbles. Furthermore, if there were a vertical gradient of grain size within the disk, we would expect to be measuring the smallest grains present, since they would have the largest scale height. For both of these reasons we consider our estimate of ~1 mm to be characteristic of the smallest grains that could be present in the disk; the bulk of the disk might consist of even larger grains.

We gratefully acknowledge the support of NASA through its Origins of Solar Systems program and the Keck Principal Investigator's Data Analysis Fund for this research. We appreciate helpful comments from Eugene Chiang, Martha Gilmore, James Greenwood and Eric Jensen.

Correspondence should be addressed to W.H. (e-mail: wherbst@wesleyan.edu).

Figure 1. Schematic drawing showing our interpretation of the geometry of the KH 15D system. Only the part of the system to the right of the solid line in the bottom panel is visible, due to the opaque screen. The elliptical orbits of stars A and B are shown, as is the occulting edge, which advances from left to right. At present, the screen is just intercepting the outer edge of the orbit of star A, so that for a while, neither star will be seen above the occulting screen. The inner disk radius is located at 0.6 AU, which is approximately the location of the 3:1 orbital resonance that defines it according to dynamical models.[14] The outer disk is located at 5 AU based on the precession time scale of ~1000 years.[5,9] The important difference between this model and others is that the reflected light is now recognized to come from evolved solids, probably of mm size, that have condensed in the terrestrial planet formation zone of this pre-planetary disk. Disks like this have long been hypothesized to exist as an intermediate step between the gas-rich, dust-laden disks seen around Classical T Tauri stars and the so-called "debris" disks seen around older stars such as β Pictoris[15] that contain dust due to fragmenting collisions between meter or kilometer-sized planetesimals. The KH 15D disk is remarkably dust free and represents an intermediate stage between a Classical T Tauri star disk and a debris disk. Although not depicted here, there is still some gas in the disk as evidenced by weak accretion signatures, including forbidden lines of sulfur and oxygen and extended emission wings to hydrogen lines.[16,17]

Figure 2. Brightness variation of KH 15D with elevation of star A. Abscissa units are radii of star A (1.3 $R_\odot \approx 10^9$ m). Dashed lines show the points where the screen first contacts the limb of the star ($\Delta=1$), where the star is half obscured ($\Delta=0$) and where it is just fully obscured ($\Delta=-1$). The inset shows the behaviour after occultation on a modified scale for clarity. The rise in brightness that occurs for elevations below -6 is due to the influence of star B, which is closer to the screen edge than star A at those orbital phases. The solid line is a "proof-of-concept" model of the light curve discussed in more detail in the Supplementary Notes. It represents a fifth-order polynomial fit to the predicted flux of the system based on this expression:

$$\text{Flux } (\Delta) = \exp(-\tau(A)) + 1.37\exp(-\tau(B)) + C_1 f_{refl}(\Delta) + C_2$$

where $f_{refl}(\Delta) = \Omega_{ring}(A) + 1.37*\Omega_{ring}(B)$ and $\Omega_{ring}$ is the solid angle subtended by the top of the disk (as a fraction of $4\pi$ sterradians) as seen from the relevant star. The luminosity of star B is 1.37 times that of star A.[5,18,19] The disk is modelled as a flat ring of inner radius 0.5 AU and outer radius 5 AU.[9,14] The exponential factors account for transmitted light, where $\tau = 1.46 - 1.46\Delta$. The parameters $C_1$ and $C_2$ were chosen to roughly fit the data by eye and have the values $C_1=2.0$ and $C_2=0.02$ for this "proof of concept" model. Physically, $C_1$ controls the relative amount of phase-dependent scattered light and depends on the geometry, albedo and phase function of the particles. $C_2$ allows for a small amount of invariable scattered light.

Figure 3. Spectral variation of KH 15D with elevation of star A. What is shown are cross-correlation functions of KH 15D spectra with a K7V synthetic spectrum over the spectral range 0.644-0.645 μ. The telescopes and instruments employed were: the 10 m Keck I Telescope and HIRES echelle spectrometer, the UV-Visual Echelle Spectrograph on the European Southern Observatory's 8.2 m Very Large Telescope, and the 6.5 m Magellan II telescope and the MIKE echelle spectrometer. The spectral region includes several photospheric lines, primarily due to Ba I, and is not affected by emission lines or terrestrial features. The vertical dashed lines indicate the radial velocity of star A (or B, for the mid-eclipse spectrum) based on our model.[5,20] The dashed-dotted lines show the expected velocity of the reflected component if the mirroring particles had the systemic velocity of 18.65 km/s and were located directly behind the star from our vantage point. When $\Delta > -1$, the light from KH 15D is dominated by star A (upper row, left three panels). When star A is fully occulted from our vantage point (middle and bottom rows), but still visible to the reflecting particles on the top side of the back of the disk  $(-6 < \Delta < -1)$, it is the reflected light from star A that contributes to the system's light and produces the red-shifted radial velocity component.  The spectrum in the upper right panel was obtained near mid-eclipse in 2004, at a time when $\Delta \equiv \Delta(A) = -7.6$ and $\Delta(B) = -7.2$. Hence, star B, which is also brighter, would be the expected source of most of the system's light in our model and its velocity and predicted reflection component are shown.

Figure 4. Colour variation of KH 15D with elevation of Star A. Error bars are ±1 s.d. The top panel shows the V-I colour of the Cousins system,[11,12] with effective wavelengths at 0.55 and 0.80 μ, respectively, while the bottom panel compares brightness in J (1.26 μ) and H (1.66 μ).[21]  In the optical range, KH 15D has a stable colour prior to the beginning of the eclipse (Δ > 1), becomes steadily bluer during eclipse (1 > Δ > -1) and maintains the bluer colour during the phases when reflected light dominates (Δ < -1). In the near-IR, there is also a transition to a bluer light during the reflection phases. The solid line shows the average colours out of eclipse. The blueness is consistent with what one might expect from unweathered silicates having particle sizes significantly larger than in the ISM.[23] Relatively blue spectral response is characteristic of some asteroidal material found in the solar system, such as Eucrites (pyroxene plus plagioclase feldspar). We further note that the shape of the albedo function is not consistent with scattering from ISM-sized grains nor is the fact that most of the scattered light clearly arrives by back-scattering, not forward scattering. Finally, note that there is no evidence of reddening of the direct light of either star at any phase; the occulting screen shows no evidence of submicron-sized dust.

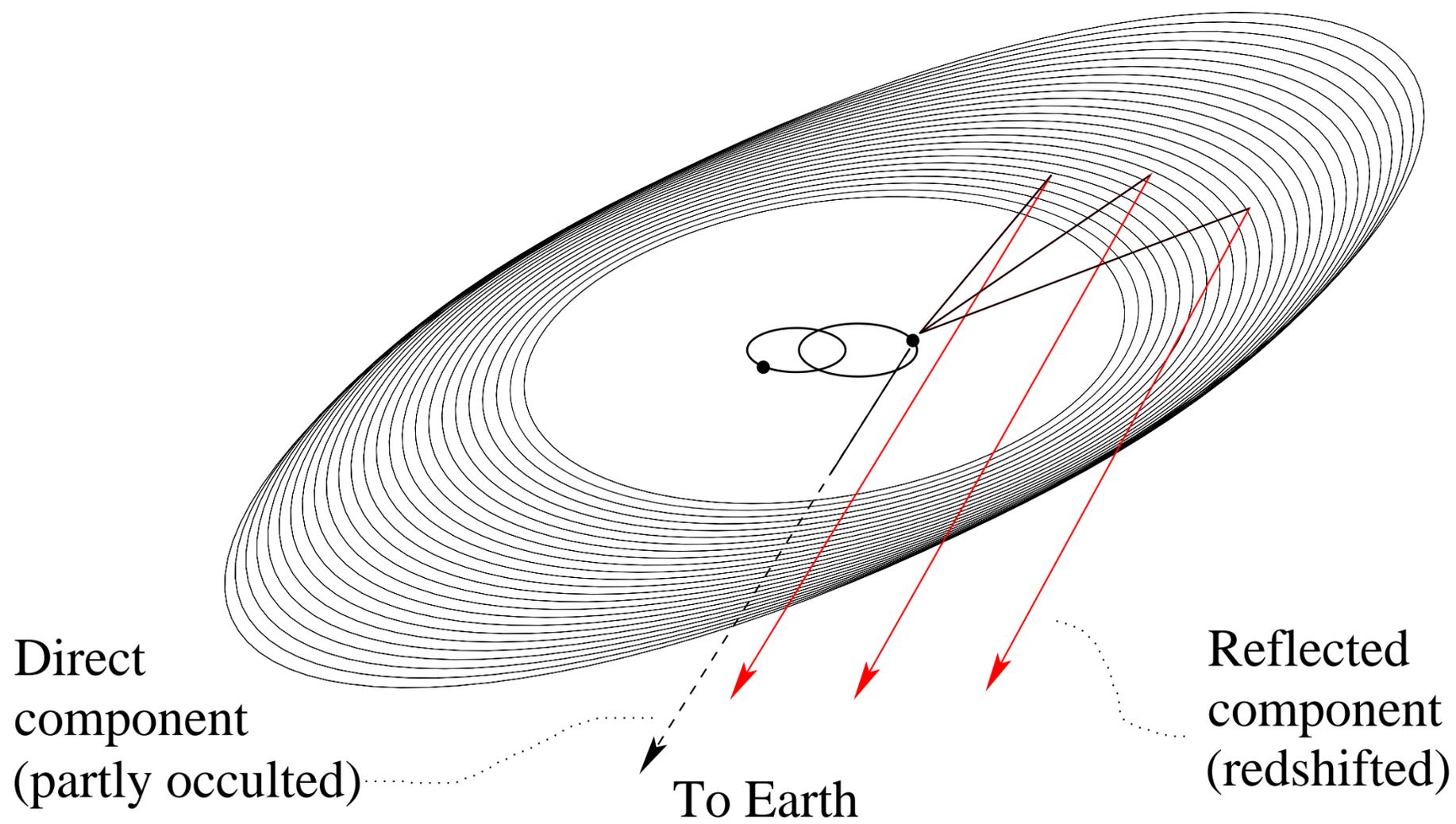

3-d view

Direct
component
(partly occulted)

To Earth

Reflected
component
(redshifted)

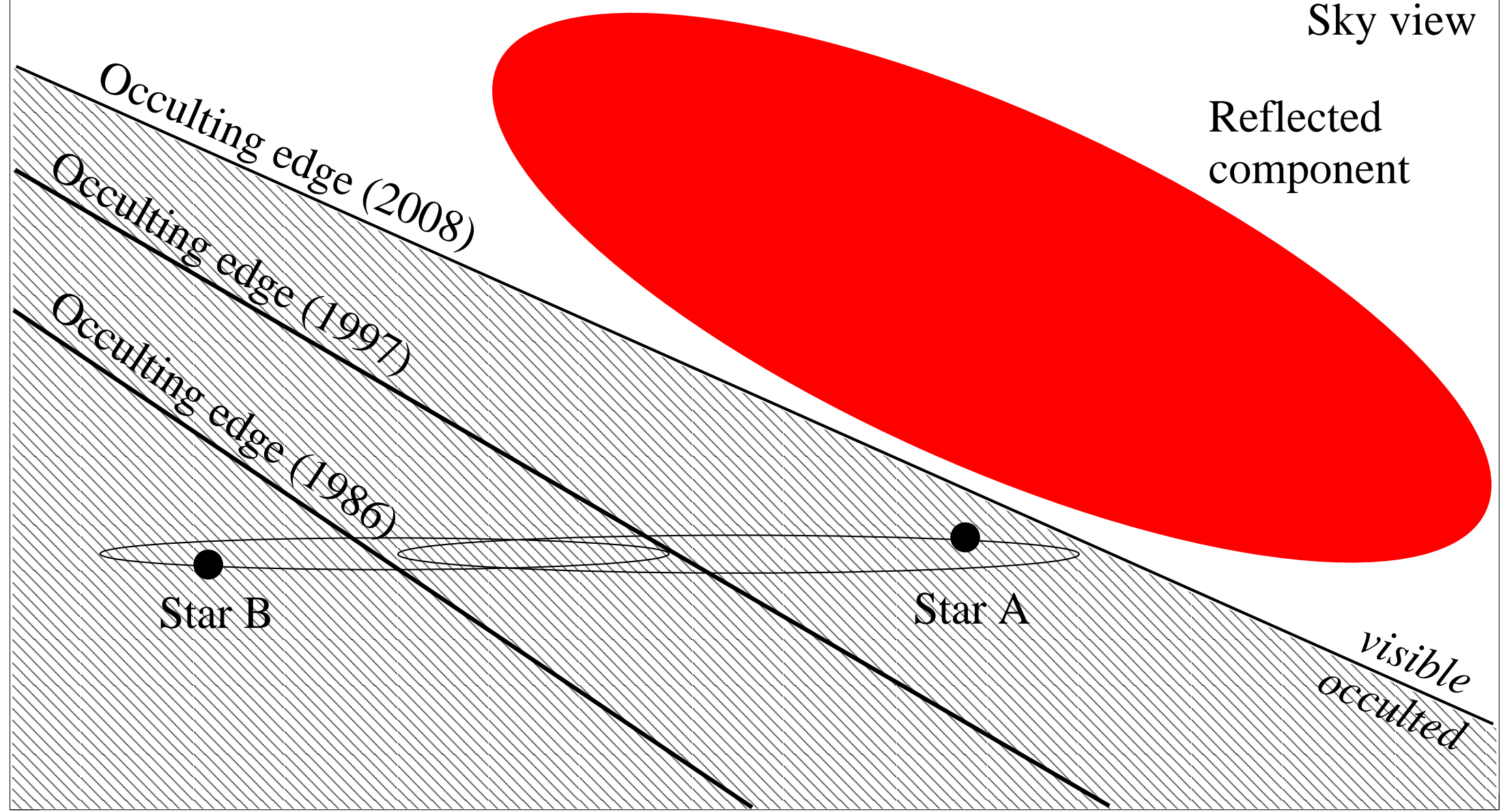

Sky view

Occulting edge (2008)

Occulting edge (1997)

Occulting edge (1986)

Reflected
component

Star B

Star A

*visible*

*occulted*

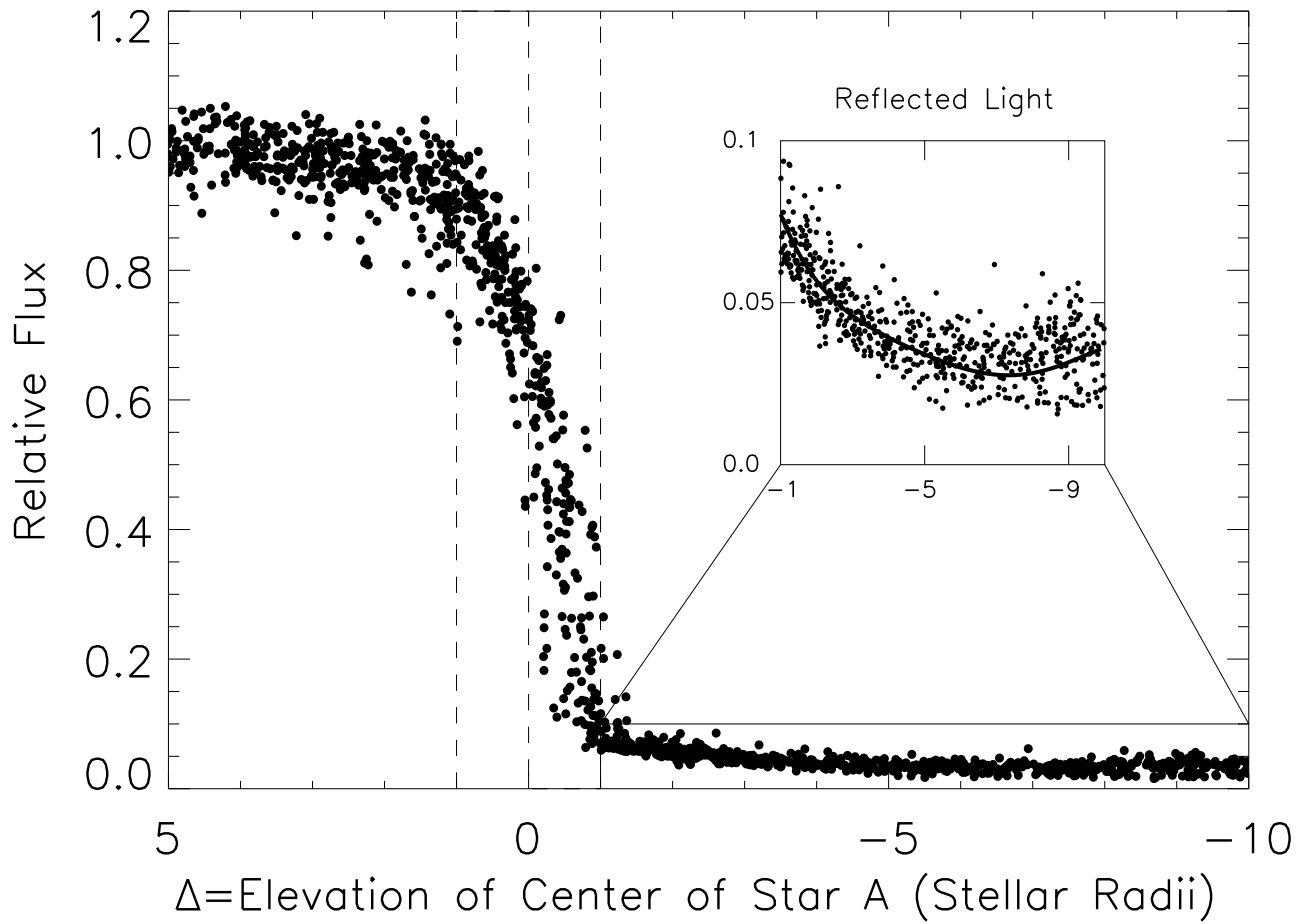

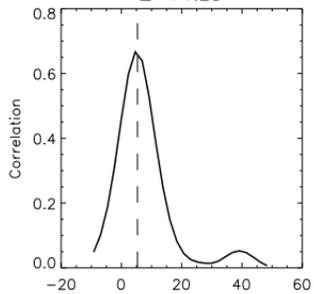
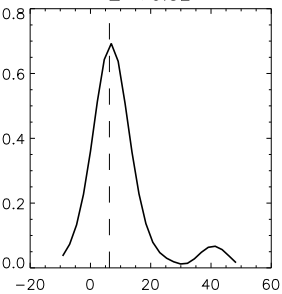
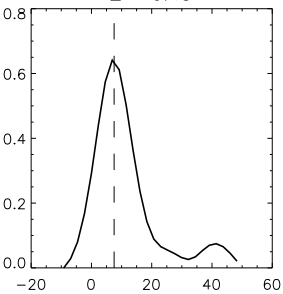
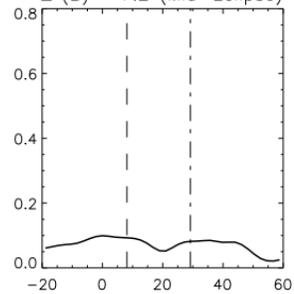
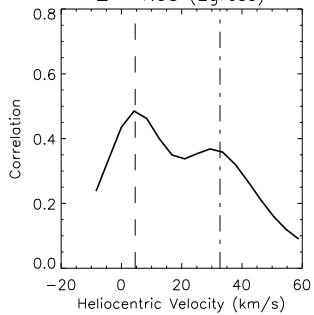
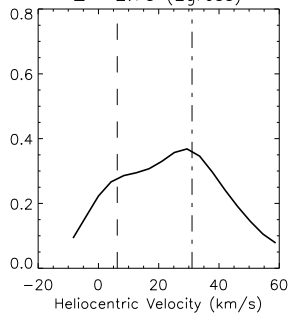
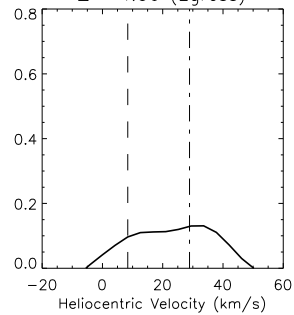
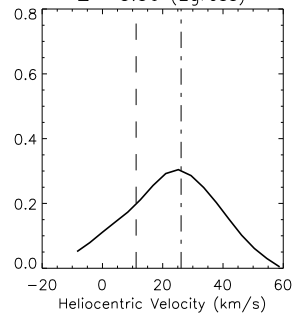
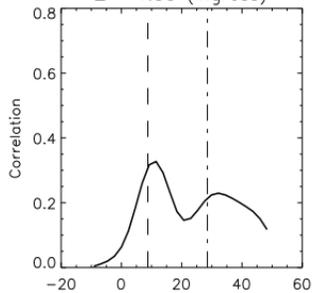
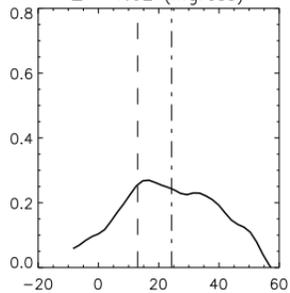
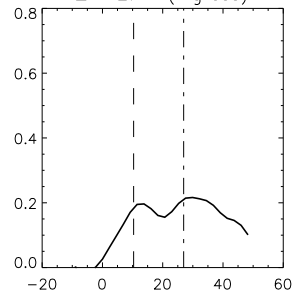
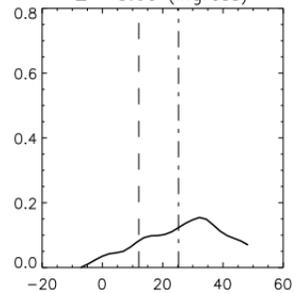

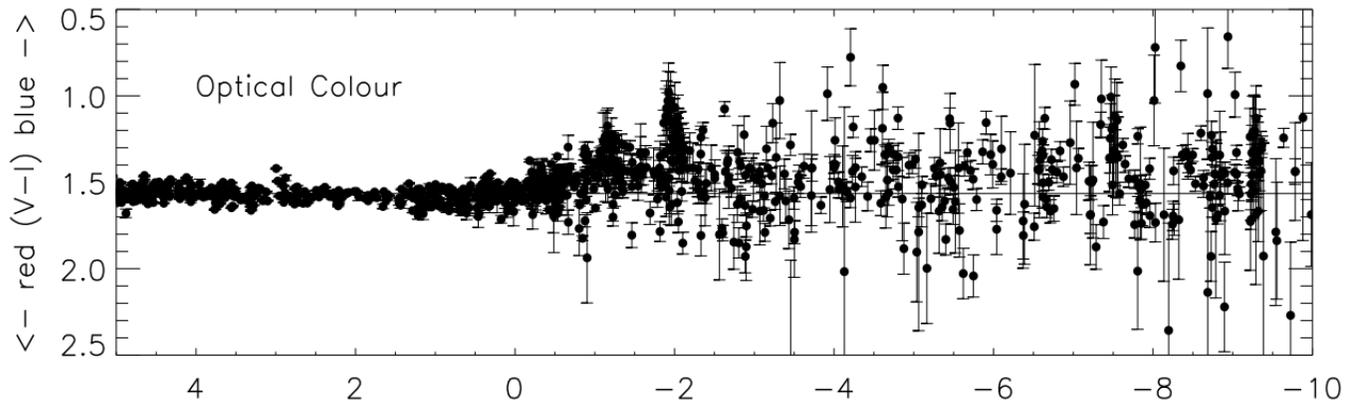

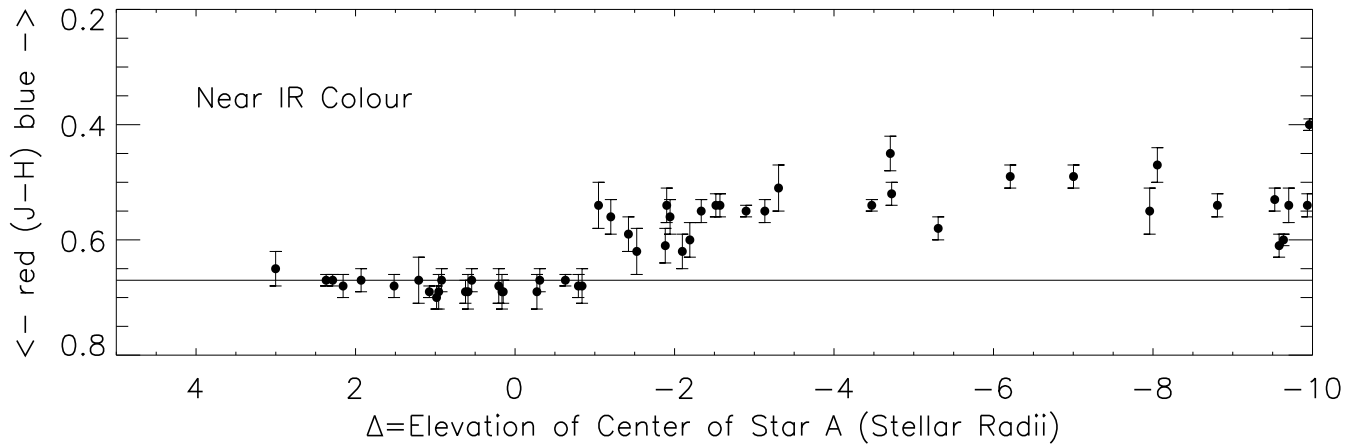

# Supplementary Notes, Equations and Figure

## 1. Variation of Optical Depth with Elevation of Star A

There are two dates for which we have good signal-to-noise spectra that clearly reveal the direct stellar component due to star A and for which we have concurrent photometry, within 0.1 days. They are JD 2453718.8 ($\Delta = -2.75$) and JD 2453719.8 ($\Delta = -1.58$). The cross-correlation functions on these nights are displayed in the second row of Figure 3. On these spectra, a two-component Gaussian fit, indicates that the fraction of light in the component matching the stellar velocity (star A) is 0.15 and 0.4, respectively. The measured I magnitude on these nights is 17.45 and 16.90, respectively. Therefore, we infer that the I magnitude of star A was $I(starA) = 17.45 - 2.5log_{10}(0.15) = 19.51$. A similar calculation on the next night yields a brightness for the direct component from star A of I = 17.89. Since the known brightness of star A when unobscured is I = 14.45, we can infer the extinction optical depth ($\tau$) at these times as $\tau = 1.086\Delta m = 1.086(19.51 - 14.45) = 5.50$ and 3.74, respectively.

There is also one date on which we can put a limit on the optical depth to star B, based on a very low signal/noise spectrum that nonetheless does resolve itself into two components, one having close to the radial velocity predicted for the star by our model (see upper right panel of Fig. 3). From this spectrum we can (crudely) estimate that no more than half of the light is coming from star B. It is likely that some, or even most, of the light in the more blue-shifted component is actually reflected light from star A. For this reason, we can only find a limit to the optical depth to star B at this phase. Nonetheless, it is an interesting limit because it is the only direct constraint we have on the optical depth to star B, at any time. Following the same procedure as in the previous paragraph, we can use the measured total light of the system at the time of this spectrum (I=18.21) to obtain the maximum brightness of star B, which is I < 19.0. The known unobscured brightness of star B is approximately I = 14.1, so we infer that $\tau > 5.3$ at this epoch, where $\Delta(starB) = -7.2$. We note that, at this epoch, star A is only slightly further below the horizon $\Delta(starA) = -7.4$ so there is a distinct possibility that we are seeing reflected light from both stars as the sources of the light and that there is no direct light visible whatsoever (i.e. $\tau$ may be extremely large for both stars, as a simple extrapolation from the two measurements for star A would suggest).

From this discussion, we conclude that $\tau(\Delta = -1) = 2.9 \pm 0.1$ and that star B is not an important source of direct light in the ingress or egress spectra shown in Fig. 3. As argued in the main journal article, the red-shifted component to the radial velocity during these phases cannot be attributed to star B without the *ad hoc* hypothesis that the extinction optical depth to the star at those phases was much less than at times when it was closer to the obscuring edge. There is no evidence for such an assumption in the present data and some evidence to the contrary, as discussed in the previous paragraph. A graphical representation of the data is shown in Supplementary Figure 1. A fit to the data yields

$$\tau(A) = 1.46 - 1.46\Delta.$$

## 2. Derivation of the Model Light Curve

We model the flux from KH 15D in the I band of the Cousins system as consisting of a direct and reflected component from each star as well as a general reflection from the entire system. The out-of-eclipse brightness of star B is known to be 1.37 times larger than star A. The total direct flux from stars A and B can, therefore, be written as:

$$f_{dir} = e^{-\tau(A)} + 1.37e^{-\tau(B)}$$

where we adopt the relation between $\tau$ and $\Delta$ given above and assume that the same relationship holds between $\tau(B)$ and $\Delta(B)$.

Our model for the reflected light component is that it is starlight back-scattered from a flat, optically thick ring of inner radius $R_{in}$ and outer radius $R_{out}$. We assume that only one side of this ring, call it the top, is visible from Earth. We further assume that the amount of reflected light from star A reaching Earth is proportional to the solid angle subtended by the top of the disk as seen from star A, and similarly for star B. If a star is located at a height, h, above a disk of radius, R, then the fraction, $\Omega$ of the whole sphere subtended by that disk is

$$\Omega = \left(\frac{1}{4}\right)\left(1 - \frac{1}{\sqrt{1 + \frac{h^2}{R^2}}}\right)$$

and the fraction subtended by a ring of inner radius, $R_{in}$ and outer radius, $R_{out}$ is

$$\Omega_{ring} = \Omega(R_{out}) - \Omega(R_{in}).$$

The total reflected flux is then calculated as:

$$f_{refl} \propto \Omega_{ring}(A) + 1.37\Omega_{ring}(B)$$

where we estimate h as the projected height of each star above (or below) the location of the center of mass of the binary, based on Model 3 of Winn et al (2006). If the star is fully above the center of mass, then it is taken to fully illuminate the top side of the disk. If it is partially above the center of mass, then its illumination of the top of the disk is proportionally reduced in the model. If the star is fully below the center of the mass then it does not contribute to the scattered light, in this model. The constant of proportionality ($C_1$) is determined by fitting the model to the data.

Combining the direct and reflected components and introducing one additional free parameter, $C_2$, that allows for some phase-independent light scattering in the vicinity of the system, one has the expression given in the text. Certainly our model for the reflected light is only a crude approximation to reality, The ring cannot actually be flat or it would not both obscure foreground light and reflect background light. It must be warped, flared and/or corrugated. One would expect that this would occur as a result of the fact that the binary orbit and the ring plane are tilted with respect to one another. In our model, the fitting parameters $C_1$ and $C_2$ obtain their values on account of these factors, as well as the albedo and phase function of the reflecting particles.

## 3. Derivation of the Expression for Optical Depth and Limit on the Size of the Grains

The optical depth, $\tau$, corresponding to a path length, L, through an obscuring medium consisting of uniform particles of number density, n, and cross-section, $\sigma$, is

$$\tau = n\sigma L.$$

Further, n may be written as N/V, where N is the total number of grains within the disk volume V, assuming a uniform distribution. For spherical particles of radius, r, and density, $\rho$, the mass of a grain is $\frac{4}{3}\pi r^3 \rho$ and the number of them is $\frac{M}{m}$ where M is the total mass of obscuring grains in V. A flat disk of radius, R, and thickness, T, has volume $V = \pi R^2 T$ so combining elements yields,

$$n = \frac{3M}{4\pi^2 r^3 \rho R^2 T}.$$

For the cross-section of a particle we may write, quite generally, that

$$\sigma = Q_{ext}\pi r^2$$

where $Q_{ext}$ is the ratio of the extinction cross section to the geometric cross section. $Q_{ext} = 2$ applies for Mie scattering when the grain radius is large compared to the wavelength of light. Mie theory also predicts strongly forward scattering phase functions so it is probably inconsistent with our result that back-scattered radiation is required to explain the spectral data. For large objects the phase function is dominated by back scattering and $Q_{ext} = 1$.

Here we shall adopt $Q_{ext} = 1$. Combining all elements, we have

$$\tau = \frac{3ML}{4\pi r \rho R^2 T}$$

If we now express M in Earth masses, r in mm, $\rho$ in gm/cm$^3$, and R in AU, the expression reduces to the one given in the main journal article, namely

$$\tau = 64 M r^{-1} \rho^{-1} R^{-2} \left(\frac{L}{T}\right)$$

From the data, we know that $\tau = 2.9$ when star A is one stellar radius below the horizon (i.e. it has just become fully occulted; see Supplementary Fig. 1). The path length, L, at that point must be of order a stellar radius, otherwise the occulting feature would be extraordinarily well aligned with our line of sight or a remarkably peaked structure. Evidence that it is neither comes also from Supplementary Figure 1, which shows that the slope of the $\tau$ vs. $\Delta$ relation is 1.46. If the occulting feature were highly peaked, then the slope would be much smaller, while if it were extremely flat, it would be much higher.

We take the disk thickness to be 20 stellar radii, assuming that the disk is not quite edge on at the present time, but is nearly so, and the projected distance from the center of mass of the system to the top of the occulting edge is about 7 stellar radii. It may be that a dynamical model of the precessing disk and binary star can provide a better limit on this, but that is beyond the scope of this paper. It is actually the ratio L/T that enters into our calculation and we assume for this order of magnitude estimate that it is 0.05.

With these values, we may rewrite the above equation for $\tau$ as an equation for r, namely

$$r \approx \frac{1}{\rho} \left(\frac{M}{R^2}\right) \; mm.$$

The expected density of grains is 1-3 gm cm$^{-3}$ where the lower value would apply if they are not well compacted or icy and the higher value for silicates. Taking $\rho = 3$ allows us to write a final approximate expression for r, namely $r \approx \Sigma$, where $\Sigma$ is the surface density of solids in the disk in units of Earth masses per square AU. If our own solar system is any guide, we might expect that the surface density of solids in the terrestrial zone will be of the order of 1. Note that the larger this value is, the larger the solids must be, in order to keep the optical depth to its measured value of 2.9 when the star is just occulted.

While there are uncertainties in all of the parameters it is clear that an order of magnitude estimate of the grain size by this method suggests r ~ 1 mm, or the size of the ubiquitous chondrules known to have populated the terrestrial zone of the Solar System prior to planet formation. This may, of course, simply be a coincidence. However, the potential opportunity to study, by reflectance spectra, a population of chondrule-sized objects in an exosolar preplanetary system should not be missed and continued precession of the disk means that the opportunity is ephemeral.

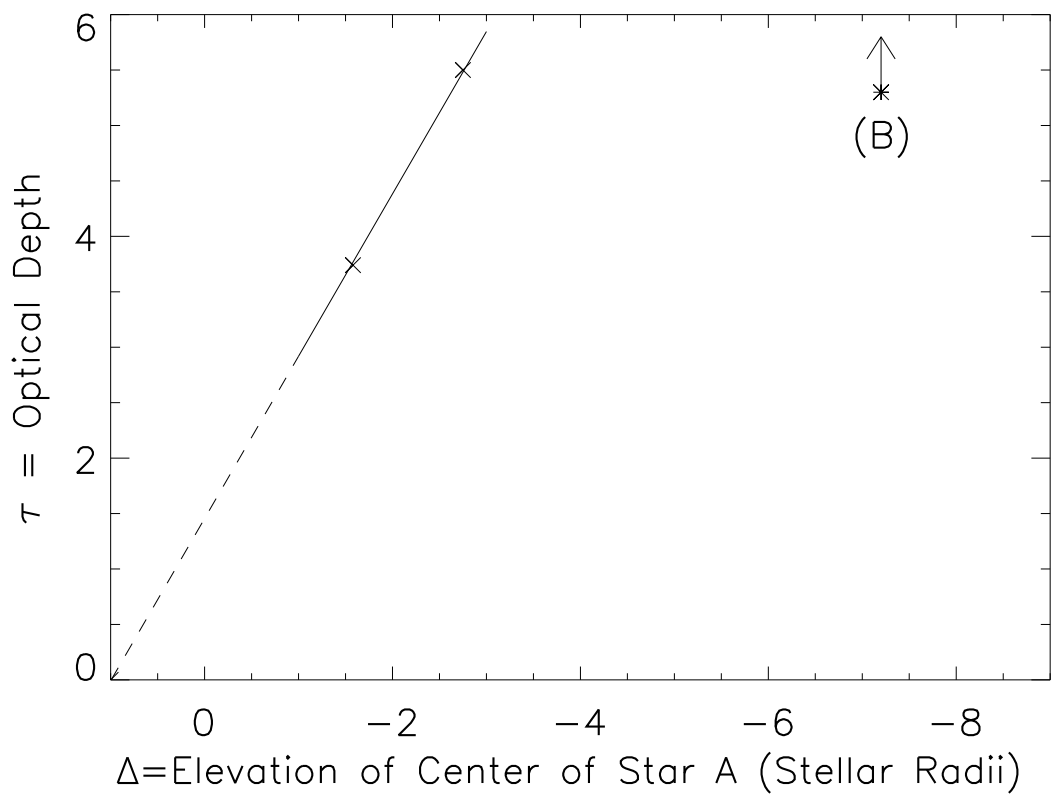

Fig. 1.— The optical depth of the transmitted light through the occulting feature can be measured at two elevations, as discussed above, by separating the total system brightness into a direct and reflected component. These measurements are plotted as X's. A third constraint is that the optical depth is zero when the star is completely above the horizon (Elevation = +1). The result of a least squares fit to these three points is shown. For $\Delta < -1$ (solid line) the star is fully occulted and it is clear that $\tau(\Delta = -1) = 2.9$. The dotted portion of the line marks elevations where parts of star A are not fully obscured. A similar calculation for star B at one epoch yields a lower limit to the optical depth to that star at a time when it was closer to the occulting edge than star A.